\documentclass[twocolumn,times]{aastex63}
\usepackage{cleveref}
\usepackage{enumerate}
\usepackage{enumitem}
\received{2021 May 07}
\revised{2021 July 08}
\accepted{2021 July 27}

\shorttitle{Periodic Variable Stars and Eclipsing Systems in LCO Photometric Data}
\shortauthors{Sanghi, A., Vanderbosch, Z., P., \& Montgomery, M., H.}

\begin{document}

\title{\textbf{Identifying Periodic Variable Stars and Eclipsing Binary Systems with Long-Term Las Cumbres Observatory Photometric Monitoring of ZTF\,J0139+5245}}

\correspondingauthor{Aniket Sanghi}
\email{asanghi01@utexas.edu}

\author[0000-0002-1838-4757]{Aniket Sanghi}
\affiliation{The University of Texas at Austin, Department of Astronomy, 2515 Speedway, C1400, Austin, TX 78712, USA}

\author[0000-0002-0853-3464]{Zachary P. Vanderbosch}
\affiliation{The University of Texas at Austin, Department of Astronomy, 2515 Speedway, C1400, Austin, TX 78712, USA}
\affiliation{McDonald Observatory, Fort Davis, TX 79734, USA}

\author[0000-0002-6748-1748]{Michael H. Montgomery}
\affiliation{The University of Texas at Austin, Department of Astronomy, 2515 Speedway, C1400, Austin, TX 78712, USA}
\affiliation{McDonald Observatory, Fort Davis, TX 79734, USA}

\begin{abstract}
We present the results of our search for variable stars using the long-term Las Cumbres Observatory (LCO) monitoring of white dwarf ZTF\,J0139+5245 with the two 1.0-m telescope nodes located at McDonald Observatory using the Sinistro imaging instrument. In this search, we find 38 variable sources, of which 27 are newly discovered or newly classified (71\%) based on comparisons with previously published catalogs, thereby increasing the number of detections in the field-of-view under consideration by a factor of $\approx$ 2.5. We find that the improved photometric precision per-exposure due to longer exposure time for LCO images combined with the greater time-sampling of LCO photometry enables us to increase the total number of detections in this field-of-view. Each LCO image covers a field-of-view of $26' \times 26'$ and observes a region close to the Galactic plane ($b = -9.4^\circ$) abundant in stars with an average stellar density of $\approx 8$ arcmin$^{-2}$. We perform aperture photometry and Fourier analysis on over 2000 stars across 1560 LCO images spanning 537 days to find 28 candidate BY Draconis variables, 3 candidate eclipsing binaries of type EA, and 7 candidate eclipsing binaries of type EW. In assigning preliminary classifications to our detections, we demonstrate the applicability of the {\em Gaia} color-magnitude diagram (CMD) as a powerful classification tool for variable star studies.
\end{abstract}


\section{Introduction}

Time-domain astronomy exploits the photometric variability of astronomical sources to probe their underlying physical mechanisms as well as details of interactions with other objects. Variable stars have profound implications in astronomy and are a key driver of research beyond the realm of time-domain astronomy. Pulsating stars such as Cepheids and RR Lyrae serve as ``standard candles" and can be used for distance determinations on the cosmic scale \citep{1912HarCi.173....1L, 2013Natur.495...76P,2019Natur.567..200P, 2018ApJ...861..126R}. Stellar surface inhomogeneities on rotational variables, such as BY Draconis variable stars, can be used to infer their rotational periods to study stellar angular momentum evolution in large samples \citep{2018A&A...616A..16L}. The orbital kinematics of binary star systems allow the determination of companion masses, which in turn can be utilized to indirectly estimate other fundamental stellar parameters \citep{2010A&ARv..18...67T}. Variable stars have also been extensively used as tracers of the structure, kinematics, chemical composition, and evolution of the Milky Way and other nearby galaxies \citep[e.g.,][]{2019Sci...365..478S, 2019NatAs...3..320C, 2016AcA....66..149J, 2017AcA....67....1J}.

The above applications are enabled by a revolutionary transformation underway in the field of astronomy --- a collective move towards the era of open-access data. Combining this openness with an improvement in the depth and efficiency of time-domain surveys, specifically, the use of wide-field CCD imagers, has led to significant advancements in the field, with the number of detected variables increasing by a factor of $\approx 2$ every year. Past surveys have generated large amounts of data used to fuel variable star discoveries and analyses. Over a period of nearly 20 years, the Optical Gravitational Lensing Experiment \citep[OGLE:][]{1994ApJ...426L..69U} has detected more than 900,000 variables in the Magellanic Clouds, the Galactic bulge, and the Galactic plane, revolutionizing the study of periodic stars and eclipsing systems. More recently, \citet{2020ApJS..249...18C} have leveraged the Zwicky Transient Facility's \citep[ZTF:][]{Masci_2018} large field-of-view and faint limiting magnitude to detect more than 700,000 variable sources in the northern sky. Further, a total of nearly 200,000 variable stars have been discovered by surveys such as the All-Sky Automated Survey for Supernovae \citep[ASAS-SN:][]{2014ApJ...788...48S}, the Catalina Sky Survey \citep[CSS:][]{2003DPS....35.3604L}, the Massive Compact Halo Object survey \citep[MACHO:][]{1993Natur.365..621A}, Pan-STARRS1 \citep{2016arXiv161205560C, 2020ApJS..251....7F, 2020ApJS..251....3M, 2020ApJS..251....6M, 2020ApJS..251....5M, 2020ApJS..251....4W}, and the {\em Gaia} mission \citep{2016A&A...595A...1G}

\begin{figure}[t]
    \centering
    \includegraphics[width=1.00\linewidth]{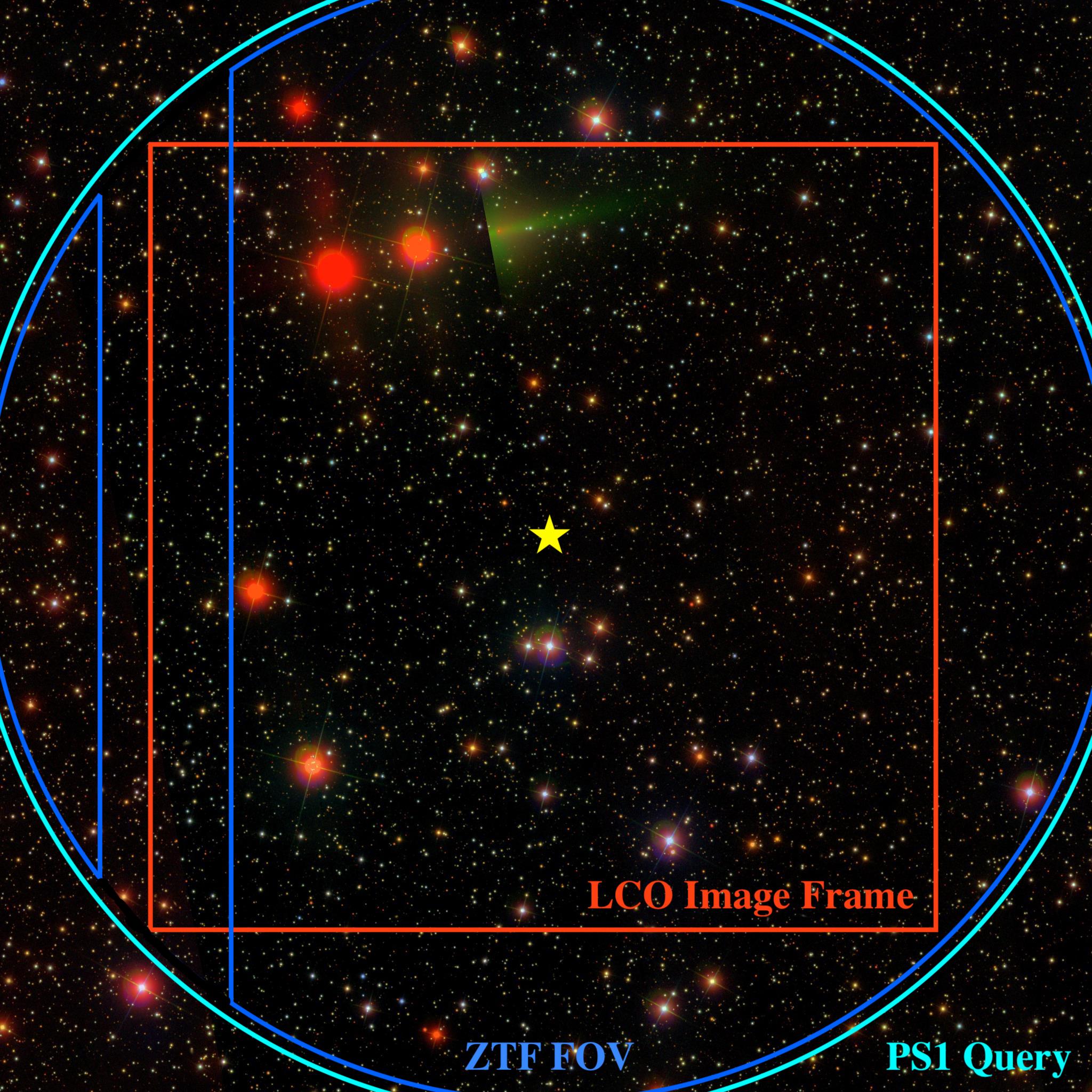}
    \caption{SDSS color image of the sky representing the coverage and field-of-view (FOV) of our LCO image frame (red) and the ZTF survey (dark blue). We also overplot the region covered by our query of the Pan-STARRS1 DR2 catalogue (cyan). The center of the LCO image frame is marked with a yellow star. Our images cover a small patch of sky that falls within a ZTF chip gap and is thus previously unexplored from the time-domain perspective beyond very sparsely sampled surveys like Pan-STARRS1.}
    \label{fig:1}
\end{figure}

Uniquely positioned in this data-driven revolution in time-domain astrophysics is the Las Cumbres Observatory (LCO). LCO utilizes a global distribution of observational facilities purpose-built to study transient and periodically variable objects at optical and near-IR wavelengths \citep{2013PASP..125.1031B} by implementing uniform instrumentation across its network and operating fully-robotically around-the-clock. It provides near continuous coverage of the night sky in both the northern and southern hemispheres, and utilizes a dynamic observation scheduling system \citep{2014SPIE.9149E..0ES} to enable extensive photometric and spectroscopic studies of transient and periodically variable sources. Consequently, LCO has had a strong impact on several fields in astronomy. It has enabled follow-up observations for exoplanet transit surveys such as the Kilodegree Extremely Little Telescope survey \citep[e.g.][]{2020AJ....160....4A, 2020MNRAS.499.3775S, 2020AJ....160..111R}, which has discovered a number of planets around bright stars. LCO is leading the detection and tracking of near Earth objects (NEOs) and asteroids \citep[e.g.][]{2017DPS....4911010L, 2021arXiv210210144L} and has also made the first optical follow-up observations to a kilonova observed by the LIGO gravitational-wave observatory \citep{2017Natur.551...64A, 2017ApJ...848L..33A, 2017ApJ...848L..32M}.

Since June 2019, the LCO network has been in use to acquire near-nightly images of ZTF J013906.17+524536.89, a white dwarf which exhibits transits likely caused by planetary debris \citep{2020ApJ...897..171V}. Each LCO image covers a field-of-view of $26' \times 26'$ and observes a region close to the Galactic plane ($b = -9.4^\circ$) abundant in stars with an average stellar density of $\approx 8$ arcmin$^{-2}$. Thus, besides their primary purpose, these observations provide a unique opportunity to leverage the high stellar density of the field to detect and classify variable sources in a relatively unexplored patch of sky from the time-domain perspective. Crucially, this observational dataset enables us to demonstrate the potential for discovery in long-term imaging observations as the astronomy community prepares for the Legacy Survey of Space and Time (LSST) to be undertaken by the Vera C. Rubin Observatory \citep{2019ApJ...873..111I}.

\begin{figure*}[t]
    \centering
    \includegraphics[width=0.95\linewidth]{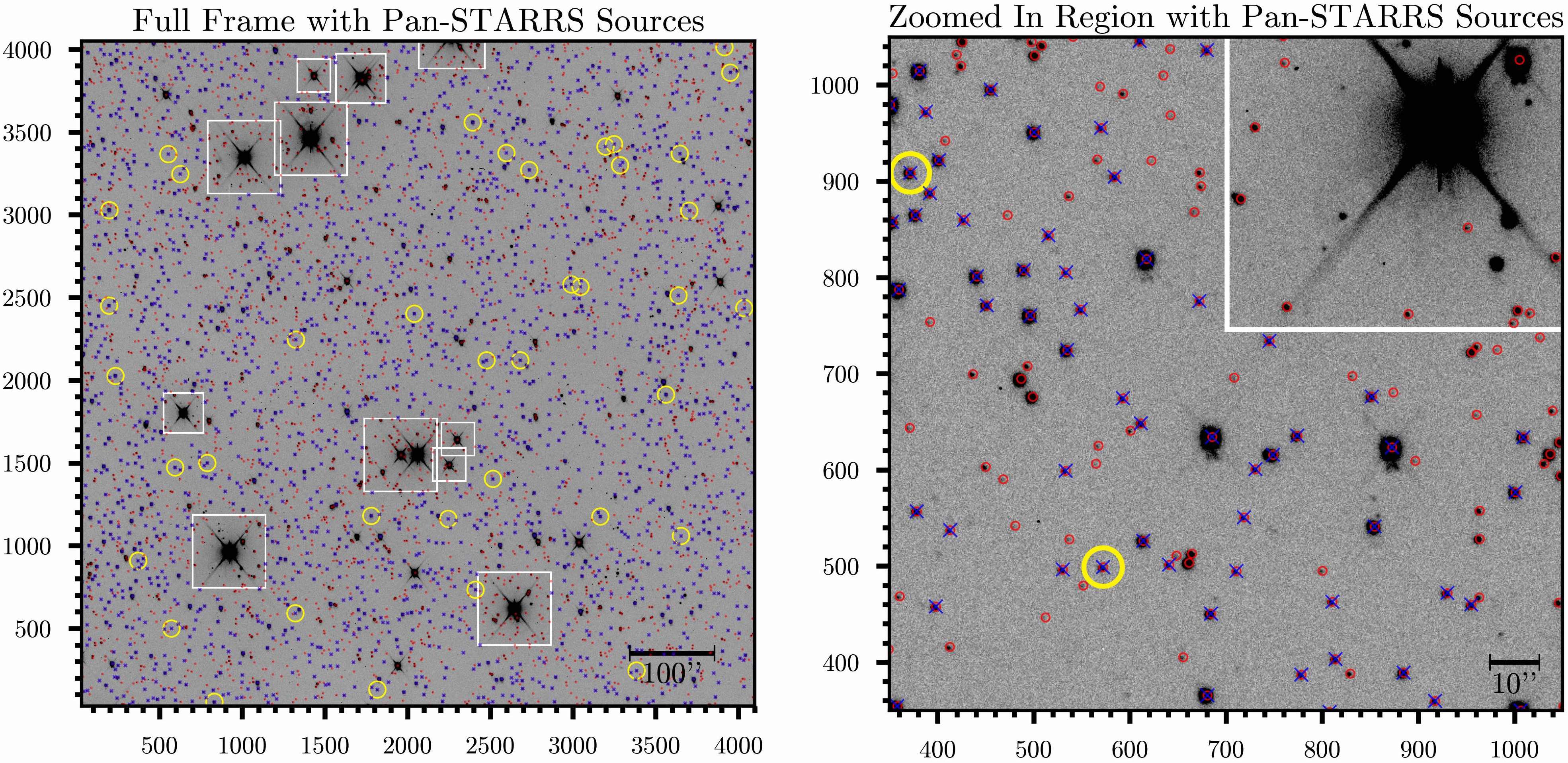}
    \caption{Plot of a full frame and zoomed in LCO image with sources (black) overlaid with the locations of Pan-STARRS1 DR2 sources (red circles). The cross-matched sources remaining after application of source cuts, on which aperture photometry is performed, are marked with blue crosses. The x and y axes are in units of pixels. The variable sources identified in this work are marked with yellow circles and the exclusion regions around bright stars are shown as white boxes.}
    \label{fig:2}
\end{figure*}

In this paper, we present the results of our search for periodic variable stars and eclipsing systems in LCO photometric data. We organize this paper into the following sections. Section \ref{methods} provides an overview of our observations, source selection and aperture photometry methods, and period analysis techniques that enable us to uncover variable stars in our dataset. Section \ref{results} details the preliminary classifications of the variable stars based on considerations of their periods, light curve morphology, and position on the {\em Gaia} EDR3 color-magnitude diagram. Section \ref{discussion} discusses and compares our results with two prominent surveys that observed in our field-of-view. Finally, we conclude in section \ref{conclusion}.

\vspace{-5pt}
\section{Methods}
\label{methods}
\subsection{Observations}
\label{2.1}
We utilize observations from the LCO 1.0-m telescope network obtained as part of an extended monitoring program for the white dwarf ZTF\,J013906.17+524536.89 \citep[hereafter ZTF\,J0139+5245,][]{2020ApJ...897..171V} which, due to its northern location and relatively faint $g=18.5$ magnitude, could only be observed by the two 1.0-m telescope nodes located at McDonald Observatory (telescope codes ELP 06 and 08). Images were acquired near-nightly using the Sinistro imaging instrument \citep{2013PASP..125.1031B} whenever ZTF\,J0139+5245 was seasonably available. For each visit, a sequence of 3$-$6 consecutive images was acquired in both the LCO $gp$ and $rp$ bands, with occasional images in the $ip$ band. Exposure times were typically between 2 and 3 minutes to achieve a signal-to-noise ratio $>30$ for the intended target, ZTF\,J0139+5245. We have a total of 1560 images spanning 537 days, with a median separation between visits of 0.98 days. Completed observations were bias, dark, and flat-field corrected via the LCO BANZAI pipeline.

\subsection{Source Selection and Aperture Photometry}
\label{2.2}
To identify good candidates for aperture photometry within our images, we first queried the second data release of the Pan-STARRS1 (PS1) survey \citep{2016arXiv161205560C} for sources within the field of view covered by an average LCO exposure. Since our images are all centered roughly on the coordinates of ZTF\,J0139+5245 ($\alpha=24.77633493\,$deg, $\delta=+52.7602692\,$deg) with a field of view of $26'\times26'$, we performed a cone search within the PS1 catalogue centered on these coordinates with a search radius of $20'$ to ensure full coverage of each LCO exposure. This query resulted in a total of 5181 sources in our image frame's field-of-view.

First, we cut this PS1 query down by excluding extremely bright stars and the objects surrounding them in boxes of width 100--220 pixels (see Figure~\ref{fig:2}) to avoid spurious detections of variability arising from scattering, diffraction, or bleeding of light from bright stars into nearby pixels in the CCD. Second, we applied magnitude limits of 12.0--19.25 to the PS1 $r$-band to exclude saturated bright objects and nearly undetectable dim objects to ensure accurate photometry. We used the $r$-band for magnitude cuts since the majority of objects in our field-of-view are red and have much fainter $g$-band magnitudes, typically between 20 and 20.5 mag for the faintest objects. This way we ensure that we can use both the $g$ and $r$-band LCO photometry for all of our objects. Third, we applied a source separation limit of $12$ arcseconds to exclude crowded objects that can cause errors in brightness measurements and centroid finding while performing aperture photometry.

\begin{figure*}[t]
    \centering
    \includegraphics[width=0.90\linewidth]{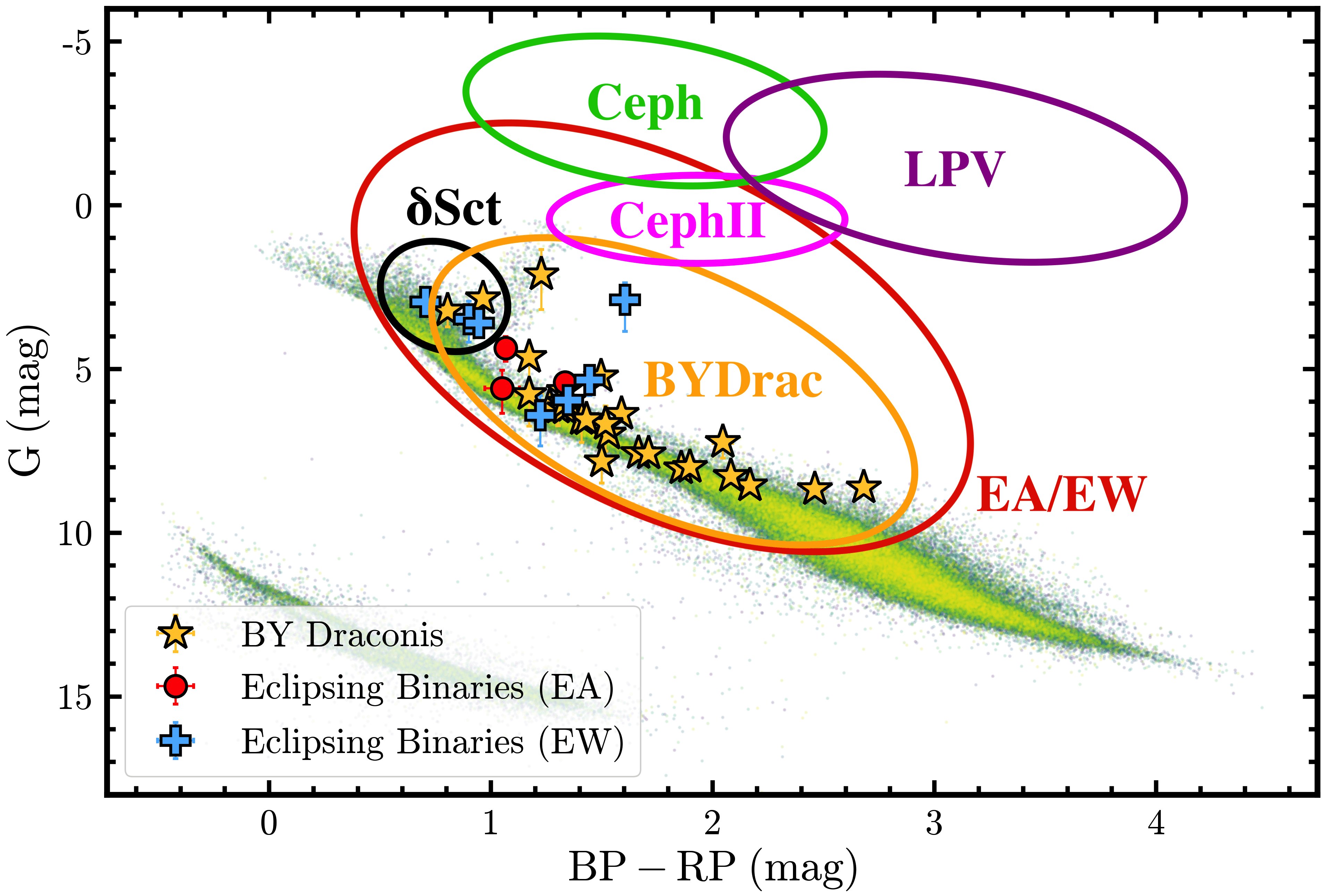}
    \caption{A $G$ vs $BP-RP$ color-magnitude diagram (CMD) depicting locations of variable sources found, for which {\em Gaia} Early Data Release 3 (EDR3) data are available. The background CMD is also constructed using the {\em Gaia} EDR3 dataset. Rough regions representing the location of common variable stars and systems in the CMD are drawn as ellipses for reference \citep{2019A&A...623A.110G}. Marker colors do not correspond to the color scheme adopted for the representation of different variability types.}
    \label{fig:3}
\end{figure*}

We reduced our sample size to 2227 sources in the photometry target list after the application of the above cuts. A total of 2954 PS1 sources were excluded from consideration of which $>50\%$ were removed on application of the magnitude cut. We cross-matched this sample with each individual image, each time excluding any objects which came within 50 pixels of the CCD edge. Sources were located by converting each source RA and Dec coordinate to pixel coordinates using the LCO image header WCS information, followed by a 2D Gaussian fit to identify the centroid. A 1D Gaussian was then fit to the radial profile of each PSF to determine the median FWHM of the image. Using the Astropy-affiliated {\sc photutils} package \citep{2013A&A...558A..33A, 2019zndo...2893252C}, we then performed circular aperture photometry on each source with an aperture radius equal to the median FWHM. Local sky subtraction was performed using a circular annulus centered on each source with inner and outer radii of 32 and 48 pixels respectively. The median sky counts per pixel within each annulus were subtracted from the respective apertures.

To calibrate our LCO photometry onto the PS1 magnitude scale, we followed the methods described in \citet{2020ApJ...897..171V}. The difference between PS1 and instrumental magnitudes ($m_d = m_{\mathrm{PS1}} - m_{\mathrm{LCO}}$) was measured while filtering for outliers and likely galaxy candidates. We then solved for a zero-point offset ($z$) and color term ($c$) with a least squares fit to $m_d = z + c\,(g_{\mathrm{PS1}} - r_{\mathrm{PS1}})$.

We also obtained additional time-series photometric data for each source (if available) by querying ZTF DR4 to improve the accuracy of our period analysis. Following recommendations in the ZTF Science Data System Explanatory Supplement\footnote{\url{http://web.ipac.caltech.edu/staff/fmasci/ztf/ztf_pipelines_deliverables.pdf}}, for objects with ZTF photometry we removed poor quality detections by requiring $\textit{catflags} = 0$ and $|\textit{sharp}| < 0.5$.

\subsection{Period Analysis of Candidate Variables}
To assess variability in our sample of objects covered by the LCO images, we first generated Lomb-Scargle periodograms \citep{1976Ap&SS..39..447L, 1982ApJ...263..835S}. We applied barycentric time corrections to all images from both LCO and ZTF using the Astropy Python package \citep{2018AJ....156..123A, 2013A&A...558A..33A}, to ensure that any stable periodicities present in our objects are accurately recovered. For each object, we used the frequency of the highest periodogram peak to also generate a phase-folded light curve. We then visually inspected both the light curves and periodograms for each object to identify promising candidates for variability. Given our small sample size ($\approx\,$2000), we visually selected objects based on qualitative features in order to maximize the true positive fraction of candidate variables in our FOV --- objects possessing well-defined, continuously variable light curves as well as objects with a clear set of peaks in their periodograms indicating an underlying period. We refer the reader to \citet{2007A&A...475.1159D, 2011ApJ...735...68K, 2014A&A...566A..43K, 2004AJ....128.2965W, 2014AJ....148...21M, 2016MNRAS.456.2260A, 2018A&A...618A..30H, 2019MNRAS.486.1907J, 2020MNRAS.491...13J, 2018AJ....156..241H} for more detailed discussions on statistical selection techniques, more suitable for survey-size samples of sources larger than what is considered in our study.

\begin{figure*}[p]
    \centering
    \includegraphics[width=0.8\textwidth]{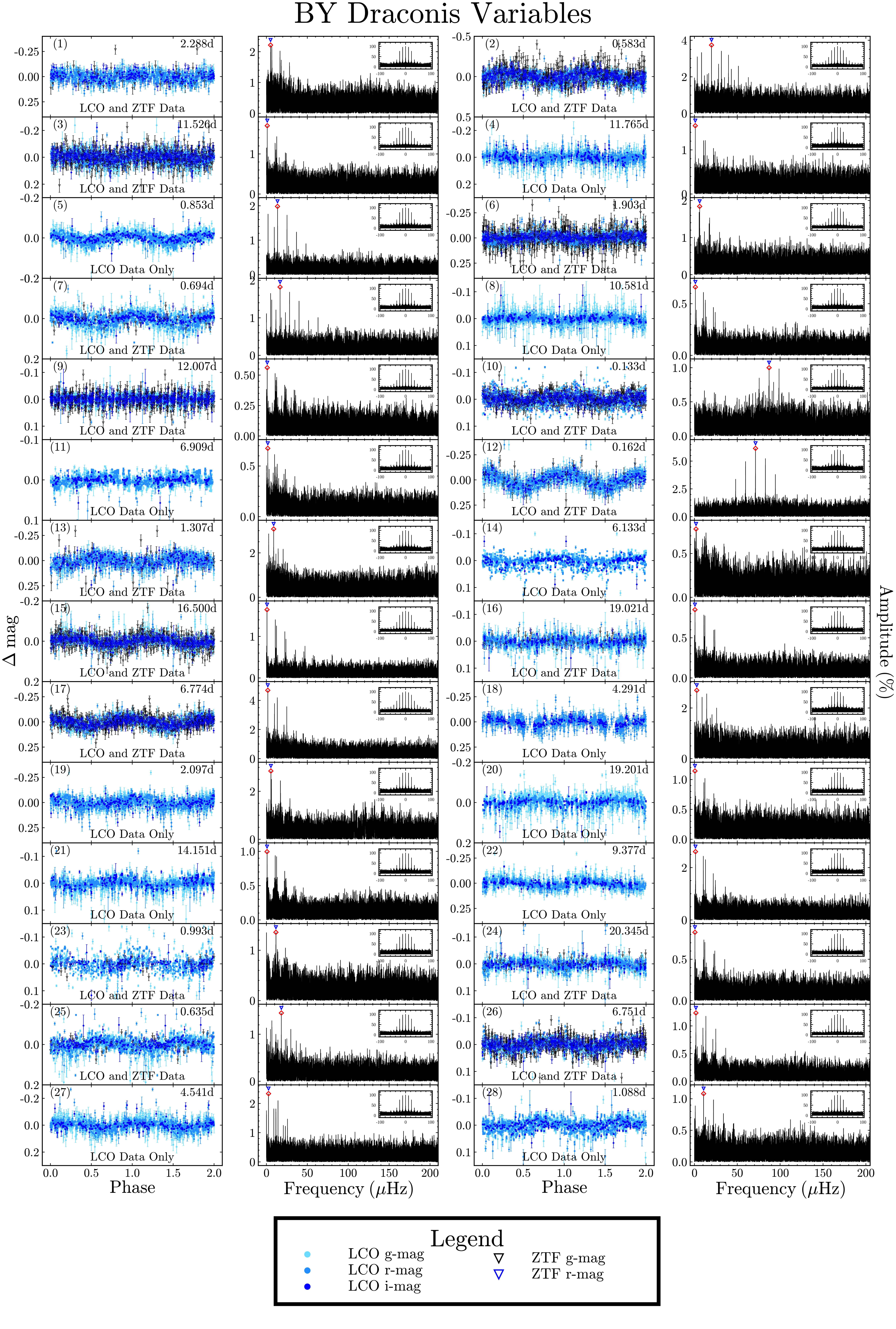}
    \caption{Light curves and periodograms for BY Draconis Variables identified in our set of variable sources. The left panel in each sub-figure represents the phase-folded light-curve data in $gp$, $rp$, and $ip$ LCO bands as well as $g$ and $r$ ZTF bands, if available, with respective error measurements. The right panel in each sub-figure represents the periodogram results with a red diamond marking the peak frequency and a blue triangle marking the folding frequency used. If the folding frequency is equal to the peak frequency, the two markers are presented one on top of another. If the folding frequency is half the peak frequency, the markers are presented in the same line, at their respective frequencies. A spectral window is displayed as an inset plot calculated based on the time-sampling of our observations. A common legend for the light curves is presented. In each light curve, the survey data and the final phase folded period (in days) used is noted.}
    \label{fig:4}
\end{figure*}

For 216 objects in our sample, we observed significant periodogram peaks recurring at near 1-day periodicities. These are usually considered to be artifacts of the window function and hence were excluded from our list of candidates. We find two possible exceptions to the above (sources 23 and 28) where the light curves indicated potential astrophysical variability. However, we note that there is an ambiguity associated with these detections due to the difficulty in distinguishing real variable sources from artifacts at near 1-day periods. In general, we identified candidate variables as objects with significant periodogram peaks which did not occur at aliases of 1-day. For these sources, we then identified the underlying period(s) using Lomb-Scargle periodograms generated with the python package {\sc Pyriod} \citep{2020AAS...23510606B}. Finally, with the refined periods from {\sc Pyriod}, we re-generated the phase-folded light curves (Figures~\ref{fig:4}, \ref{fig:5}) for each candidate variable star.

We make a couple of cautionary comments. Using the above period determination technique, the most significant peaks in the periodograms of eclipsing binaries will sometimes correspond to half the orbital period of the system if ellipsoidal variations or two equally-sized primary and secondary transits are present. Accurate period determination for binaries may thus involve setting the folding frequency to half the peak frequency obtained from the periodograms. For sources whose variability may be caused by pulsations, rotation, chromospheric activity, etc., it is unclear whether the folding frequency should be modified, as above, to determine the period. This highlights the fundamental limitation in determining the period of objects without a classification. For such objects in our dataset, wherever necessary, we present the light curve generated with no changes to the folding frequency, i.e., by setting the folding frequency equal to the peak frequency in the periodogram. We also apply a $5\sigma$ limit to the data points in each light curve using the {\sc sigma\_clip} function from {\sc astropy.stats} to exclude outliers that could increase the noise levels and potentially mask true peaks in the Lomb-Scargle periodogram. We then note that the periodograms in Figures~\ref{fig:4}, \ref{fig:5} are generated after the application of sigma clipping but the corresponding light curves still present the clipped data points for reference.

\subsection{Gaia Early Data Release 3 (EDR3) Color-Magnitude Diagram}
We obtain photometry for our candidate variable sources from the publicly accessible {\em Gaia} EDR3 dataset \citep{2021A&A...649A...1G}. We note that the large fractional parallax uncertainties precluded a simple inversion of the parallax measurement provided by {\em Gaia} to calculate the distance to each of the variable sources. Instead, we determined the absolute $G$ band mean magnitude using the geometric distance estimates provided by \citet{2021AJ....161..147B}. Errors on the relevant quantities quoted in Table \ref{table} are then derived using standard propagation of errors.

\section{Results}
\label{results}
\begin{figure*}[!t]
    \centering
    \includegraphics[width=1.0\textwidth]{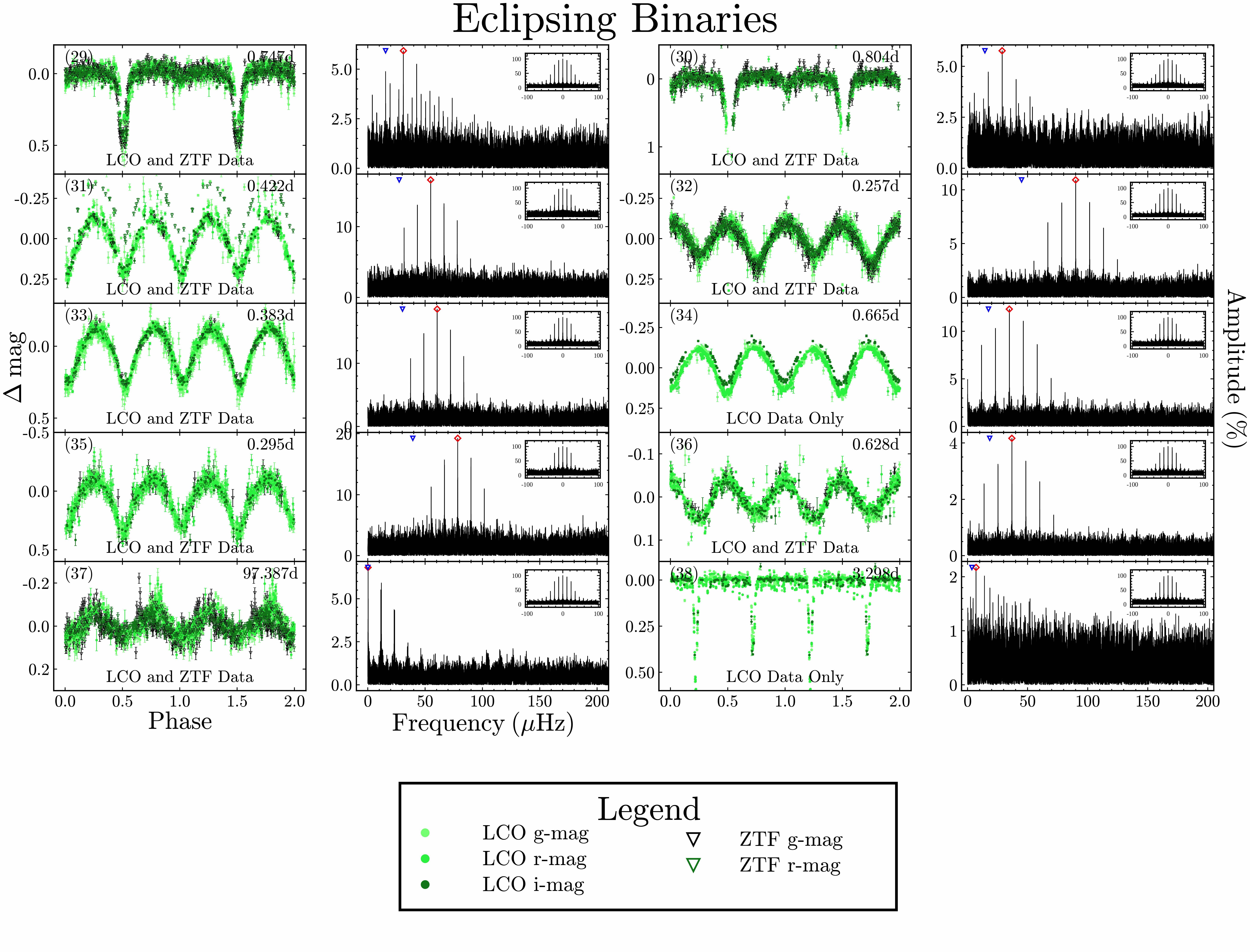}
    \caption{The same as Figure~\ref{fig:4} but for Eclipsing Binaries.}
    \label{fig:5}
\end{figure*}

We present the 38 variable sources found by visual inspection of light curves and periodograms obtained from aperture photometry of sources cross-matched with the PS1 Survey, after application of source data cuts, in Table \ref{table} along with their preliminary classifications. The phase-folded light curves and periodogram results are shown in Figures~\ref{fig:4} and \ref{fig:5}. The primary source of variability observed is due to rotation and eclipses. We describe certain general characteristics of the sources found, grouped by their preliminary classification, below.

\subsection{Variability due to eclipses and rotation}
\begin{enumerate}[leftmargin=*]

    \item \textbf{BY Draconis Variables:} These are main-sequence stars with late spectral types (K and M) that exhibit quasi-periodic light curves due to spots and chromospheric activity. The fluctuations are low amplitude, generally less than 0.5 mag. Their periods range from a fraction of a day to nearly 120 days. We classify sources (1)--(28) by this definition. The periods of our candidates are in the range 0.1--20d. This classification is supported by the location of these sources on the {\em Gaia} EDR3 color-magnitude diagram (G3CMD) along the main sequence. Since nearly every K and M star has a light curve characteristic of a BY Draconis, this category of variables is expected to have the largest number of detections in our sample, as we do indeed find. The light curves and corresponding periodograms of all BY Draconis variables are presented in  Figure~\ref{fig:4}. We note that sources (23) and (28) have a period very close to a 1 day period characteristic of artifacts in the window function. Thus, it is possible that while the light curves of these sources indicate potential variability, these might not be true BY Draconis variables. This highlights our limitation of confirming true variables with periods near the 1 day period.
    \\
    \vspace{-9pt}
    \item \textbf{Eclipsing Binaries of type EA:} These are binaries with spherical or slightly ellipsoidal components with well-separated, nearly constant light curves in between minima. Secondary minima can be absent in these types. The prototype for this class is Algol. We classify sources (29)--(31) by this definition. This classification is supported by the locations of these sources on the G3CMD along the main sequence. The periods of our candidates are in the range 0.4--0.8d. The light curves and corresponding periodograms of all eclipsing binaries are presented in Figure~\ref{fig:5}.

    \item \textbf{Eclipsing Binaries of type EW:} These are binaries where the components are nearly or actually in contact and both minima are virtually equally strong. Onsets and ends of minima are not well defined in the light curves of these objects. Their periods are generally less than one day. The prototype for this class is W Ursae Majoris. We classify sources (32)--(38) using this definition. This classification is supported by the location of these sources on the {\em Gaia} EDR3 color-magnitude diagram (G3CMD) along the main sequence, similar to type EA eclipsing binaries. The periods of our candidates are in the range 0.25--98d.

\end{enumerate}

\subsection{Variability due to pulsations}
We note that the light curves of BY Draconis variables, as presented in Figure~\ref{fig:4}, are often similar to first overtone Cepheids, and in the case of low period variables, similar to $\delta$ Scuti stars, leading to an ambiguity in assigning a preliminary classification. \citet{2019A&A...623A.110G} find constrained regions in a $G$ vs $BP - RP$ {\em Gaia} color-magnitude diagram for each type of variable star. We can then eliminate ambiguity in our classifications using the location of our objects in the CMD presented in Figure~\ref{fig:3}. We observe that none of our sources lie in the regions dominated by classical cepheids (Ceph), type II cephieds (CephII), or long period variables (LPV). There are a few sources that fall within the $\delta$ Scuti ($\delta$Sct) region, but these either possess transits characteristic of eclipsing binary systems or periods outside the range of periods expected for $\delta$ Scuti variables. Moreover, since K and M dwarfs are magnetically active, a certain amount of scatter and irregularity in the shape of the light curve is expected compared to the smoother light curves of Cepheids and $\delta$ Scutis. This allows us to further strengthen our preliminary classification in certain cases. Lastly, we note that source (16) is not presented in the {\em Gaia} CMD due to the lack of a parallax measurement and hence holds a larger uncertainty with respect to its preliminary classification.

\startlongtable
\centerwidetable
\begin{deluxetable*}{lllllllllll}
\vspace{-0.5cm}
\centering
\tablecaption{Variable Sources Found in LCO Images}

\tablenum{1}

\tablehead{\colhead{Index} & \colhead{Pan-STARRS ID} & \colhead{R.A. (J2000)} & \colhead{Decl. (J2000)} & \colhead{Period} & \colhead{$\langle g \rangle$} & \colhead{$\langle r \rangle$} & \colhead{$\langle i \rangle$} & \colhead{Classification} & \colhead{$G$} & \colhead{$BP - RP$}\\
\colhead{} & \colhead{} & \colhead{(deg)} & \colhead{(deg)} & \colhead{(days)} & \colhead{(mag)} & \colhead{(mag)} & \colhead{(mag)} & \colhead{} & \colhead{(mag)} & \colhead{(mag)}}

\startdata
(1) & 171470250272081779 & 25.02717 & 52.89257 & 2.288 & 19.15 & 18.35 & 17.96 & BYDrac & $7.83_{-0.37 }^{+0.66}$ & $1.50 \pm 0.04$ \\
(2) & 171180244883024042 & 24.48828 & 52.65282 & 0.583 & 19.51 & 18.33 & 17.65 & BYDrac & $7.23_{-0.39 }^{+0.49}$ & $2.05 \pm 0.04$ \\
(3) & 171340249027671181 & 24.90277 & 52.78374 & 11.526 & 18.82 & 18.07 & 17.74 & BYDrac & $6.19_{-0.48 }^{+0.55}$ & $1.30 \pm 0.02$ \\
(4) & 171110250391044651 & 25.03911 & 52.59500 & 11.765 & 19.10 & 17.82 & 17.21 & BYDrac & $8.04_{-0.21 }^{+0.20}$ & $1.86 \pm 0.03$ \\
(5) & 171140247106853549 & 24.71066 & 52.61904 & 0.853 & 17.08 & 16.11 & 15.64 & BYDrac & $6.38_{-0.10 }^{+0.09}$ & $1.59 \pm 0.01$ \\
(6) & 171120249053056178 & 24.90530 & 52.60457 & 1.903 & 19.45 & 18.04 & 17.05 & BYDrac & $8.68_{-0.16 }^{+0.17}$ & $2.46 \pm 0.04$ \\
(7) & 171490245651510736 & 24.56514 & 52.90840 & 0.694 & 17.29 & 16.09 & 15.45 & BYDrac & $7.99_{-0.04 }^{+0.03}$ & $1.90 \pm 0.01$ \\
(8) & 171230246907620573 & 24.69076 & 52.69156 & 10.581 & 17.24 & 16.27 & 15.86 & BYDrac & $6.98_{-0.07 }^{+0.09}$ & $1.53 \pm 0.01$ \\
(9) & 171360247742360866 & 24.77424 & 52.80015 & 12.007 & 16.53 & 15.78 & 15.46 & BYDrac & $6.14_{-0.09 }^{+0.08}$ & $1.27 \pm 0.01$ \\
(10) & 171290245026655286 & 24.50266 & 52.74551 & 0.133 & 15.39 & 14.92 & 14.73 & BYDrac & $2.85_{-0.23 }^{+0.24}$ & $0.97 \pm 0.01$ \\
(11) & 171200248226272288 & 24.82264 & 52.66799 & 6.909 & 15.67 & 14.90 & 14.56 & BYDrac & $5.68_{-0.09 }^{+0.09}$ & $1.33 \pm 0.01$ \\
(12) & 171480246728186487 & 24.67282 & 52.90483 & 0.162 & 19.32 & 18.74 & 18.44 & BYDrac & $5.77_{-0.89 }^{+0.97}$ & $1.17 \pm 0.06$ \\
(13) & 171490245557812506 & 24.55579 & 52.90988 & 1.307 & 19.39 & 18.53 & 18.14 & BYDrac & $6.69_{-0.56 }^{+0.82}$ & $1.52 \pm 0.04$ \\
(14) & 171310250983433394 & 25.09835 & 52.76059 & 6.133 & 15.31 & 14.61 & 14.30 & BYDrac & $2.12_{-0.77 }^{+1.06}$ & $1.23 \pm 0.01$ \\
(15) & 171320246965463740 & 24.69654 & 52.76920 & 16.500 & 17.48 & 16.39 & 15.90 & BYDrac & $7.58_{-0.06 }^{+0.08}$ & $1.67 \pm 0.01$ \\
(16) & 171240249993444675\tablenotemark{a} & 24.99945 & 52.70332 & 19.021 & 16.93 & 16.13 & 15.79 & BYDrac & \nodata & \nodata \\
(17) & 171360244167352479 & 24.41674 & 52.80152 & 6.774 & 18.59 & 17.45 & 16.93 & BYDrac & $7.58_{-0.18 }^{+0.19}$ & $1.71 \pm 0.01$ \\
(18) & 171470245496025935 & 24.54960 & 52.89607 & 4.291 & 20.25 & 18.87 & 17.57 & BYDrac & $8.63_{-0.24 }^{+0.22}$ & $2.68 \pm 0.04$ \\
(19) & 171190247395279443 & 24.73952 & 52.66563 & 2.097 & 18.93 & 18.13 & 17.73 & BYDrac & $6.54_{-0.44 }^{+0.71}$ & $1.41 \pm 0.03$ \\
(20) & 171160250745668048 & 25.07457 & 52.63950 & 19.201 & 17.56 & 16.22 & 15.35 & BYDrac & $8.56_{-0.03 }^{+0.03}$ & $2.17 \pm 0.01$ \\
(21) & 171380246037492681 & 24.60373 & 52.81834 & 14.151 & 17.02 & 16.22 & 15.88 & BYDrac & $5.99_{-0.12 }^{+0.16}$ & $1.35 \pm 0.01$ \\
(22) & 171430244750309507 & 24.47503 & 52.86571 & 9.377 & 18.81 & 17.46 & 16.72 & BYDrac & $8.27_{-0.11 }^{+0.14}$ & $2.08 \pm 0.02$ \\
(23) & 171050249931526875\tablenotemark{b} & 24.99314 & 52.54684 & 0.993 & 15.75 & 15.10 & 14.82 & BYDrac & $4.64_{-0.08 }^{+0.08}$ & $1.17 \pm 0.01$ \\
(24) & 171470246482432949 & 24.64823 & 52.89356 & 20.345 & 16.67 & 15.78 & 15.41 & BYDrac & $6.55_{-0.06 }^{+0.05}$ & $1.43 \pm 0.01$ \\
(25) & 171540244285547663 & 24.42854 & 52.95582 & 0.635 & 17.08 & 16.19 & 15.76 & BYDrac & $5.23_{-0.14 }^{+0.17}$ & $1.50 \pm 0.01$ \\
(26) & 171200245752030093 & 24.57520 & 52.66616 & 6.751 & 16.15 & 15.37 & 15.04 & BYDrac & $6.21_{-0.04 }^{+0.04}$ & $1.33 \pm 0.01$ \\
(27) & 171380245939520760 & 24.59393 & 52.81674 & 4.541 & 18.43 & 17.65 & 17.30 & BYDrac & $6.04_{-0.32 }^{+0.40}$ & $1.33 \pm 0.02$ \\
(28) & 171360251049928675\tablenotemark{b} & 25.10500 & 52.80664 & 1.088 & 16.33 & 15.97 & 15.84 & BYDrac & $3.22_{-0.30 }^{+0.49}$ & $0.81 \pm 0.01$ \\
(29) & 171320246599753528 & 24.65998 & 52.76903 & 0.747 & 17.46 & 16.73 & 16.34 & EA & $5.41_{-0.23 }^{+0.28}$ & $1.34 \pm 0.02$ \\
(30) & 171370244878453108 & 24.48785 & 52.81034 & 0.804 & 19.41 & 18.86 & 18.58 & EA & $5.59_{-0.55 }^{+0.76}$ & $1.05 \pm 0.08$ \\
(31) & 171560244345567723 & 24.43454 & 52.97254 & 0.422 & 17.06 & 16.54 & 16.32 & EA & $4.36_{-0.34 }^{+0.41}$ & $1.07 \pm 0.04$ \\
(32) & 171060248174965790 & 24.81747 & 52.55424 & 0.257 & 17.72 & 16.82 & 16.41 & EW & $5.34_{-0.26 }^{+0.47}$ & $1.45 \pm 0.02$ \\
(33) & 171480244845204434 & 24.48451 & 52.90315 & 0.383 & 17.74 & 16.98 & 16.62 & EW & $5.95_{-0.32 }^{+0.42}$ & $1.35 \pm 0.04$ \\
(34) & 171240250341981190 & 25.03427 & 52.70046 & 0.665 & 16.55 & 16.10 & 15.99 & EW & $3.49_{-0.56 }^{+0.69}$ & $0.90 \pm 0.03$ \\
(35) & 171480250400887540 & 25.04008 & 52.90570 & 0.295 & 19.43 & 18.76 & 18.46 & EW & $6.42_{-0.61 }^{+0.93}$ & $1.22 \pm 0.05$ \\
(36) & 171510247091430730 & 24.70913 & 52.92505 & 0.628 & 15.07 & 14.80 & 14.71 & EW & $2.96_{-0.16 }^{+0.13}$ & $0.71 \pm 0.01$ \\
(37) & 171070245379218746 & 24.53793 & 52.56505 & 97.387 & 17.65 & 16.71 & 16.25 & EW & $2.89_{-0.52 }^{+0.96}$ & $1.60 \pm 0.02$ \\
(38) & 171440251042963672 & 25.10429 & 52.86914 & 3.298 & 14.90 & 14.43 & 14.23 & EW & $3.60_{-0.08 }^{+0.08}$ & $0.95 \pm 0.01$ \\
\enddata

\tablenotetext{a}{{\em Gaia} EDR3 does not quote a parallax measurement.}
\tablenotetext{b}{Possible artifacts of the window function with a 1 day period.}

\tablecomments{EA and EW represent sub-types of Eclipsing Binaries and BYDrac represents a BY Draconis Variable.}

\label{table}
\end{deluxetable*}

\begin{figure*}[!ht]
    \centering
    \includegraphics[width=\textwidth]{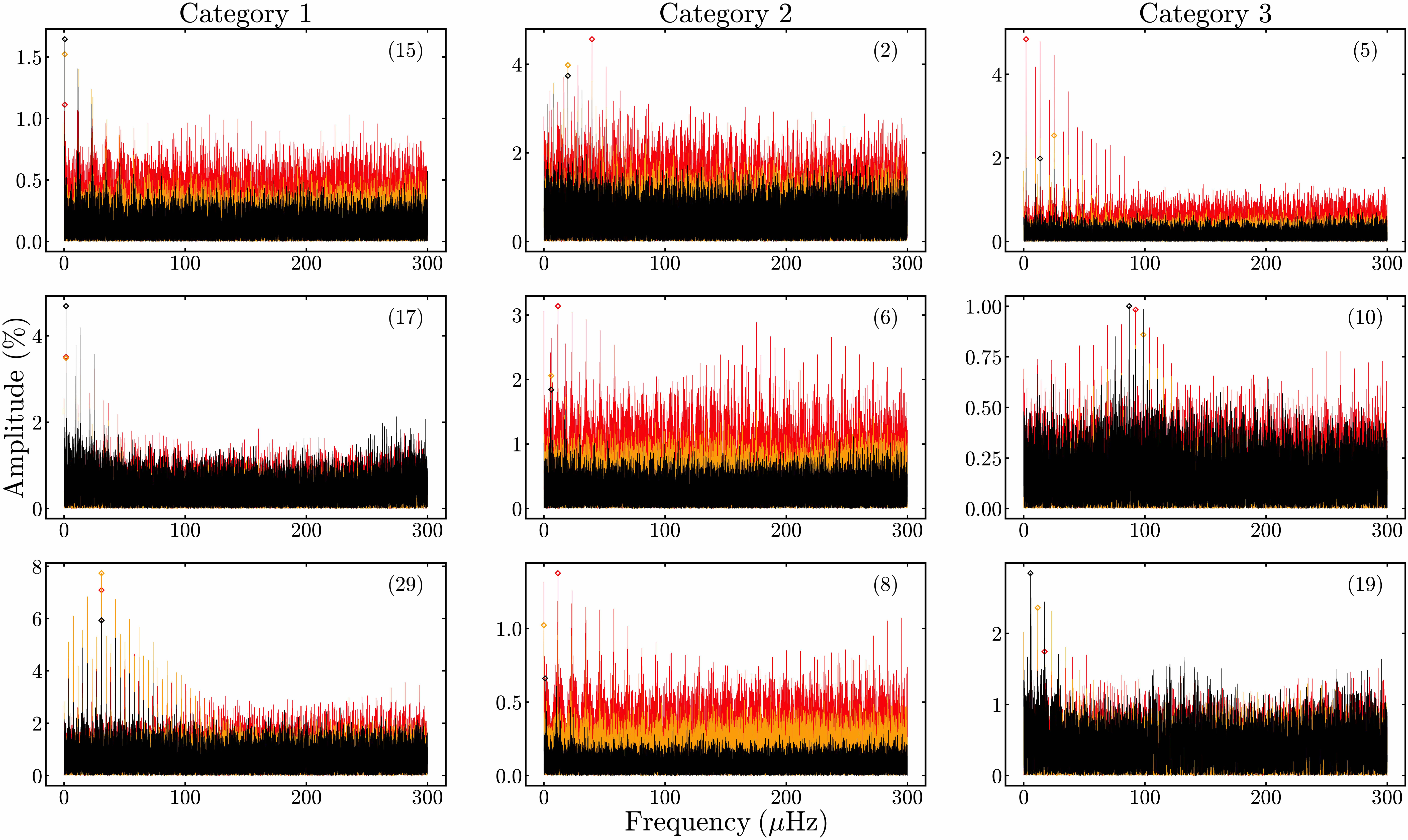}
    \caption{A comparison of the periodograms generated for each of our variable stars using three sets of survey data --- ZTF DR2 (red), ZTF DR5 (yellow), and LCO (black). The diamond markers represent the peak frequency obtained with the periodogram generated for each dataset. The first column of the panel comprising sources (15), (17), (29) represents examples of sources for which the peak frequency obtained agrees across all three datasets. The second column of the panel comprising sources (2), (6), (8) represents examples of sources for which the peak frequency obtained with LCO agrees only with that obtained with ZTF DR5. The third column of the panel comprising sources (5), (10), (19) represents examples of sources for which the peak frequency obtained with LCO does not agree with ZTF DR2 and DR5.}
    \label{fig:6}
\end{figure*}

\section{Discussion}
\label{discussion}
\subsection{ZTF Catalog of Variable Stars}
We performed a box search of size 25 arcminutes on our FOV centered on the object J013906.17+524536.8 in the catalog of ZTF variable stars provided by \citet[hereafter C20]{2020ApJS..249...18C}.
We then cross-matched our list of candidate variable stars with the results of the box search. In our image's FOV, C20 have identified 7 variable sources, i.e., approximately a factor of five less than our identification of 38 variable sources. Out of these seven sources, our cross-match yielded four matching identifications: (5), (30), (32), (37). This suggests that both sets of data are missing variable stars in this FOV. Further investigation revealed that our crowding-limitation selection criterion eliminated 3/7 sources in C20's coverage of our FOV.

C20 classify (5) as a BY Draconis variable with period 0.8556985d, (30) as an EA-type eclipsing binary system with period 0.8038496d, (32) as an EW-type eclipsing binary system with period 0.257176d, and (37) as a Semiregular variable with period 50.2286076d. Our classifications and derived periods for these objects are in agreement with the above as presented in Table \ref{table}, with the exception of (37) which we identify as an EW type eclipsing binary system with a folded period of 97.387d.

We generated periodograms for all our candidate variable objects with three sets of data --- ZTF DR2, ZTF DR5, and LCO --- to investigate the lack of variable objects identified with ZTF.  We find that there is either insufficient or no ZTF DR2 data available for 19 sources. C20 utilize ZTF DR2 and hence would not have found these sources. DR5 improves the availability of data over DR2, with only 6 sources having either insufficient or no data. We present a subset of the generated periodograms in Figure~\ref{fig:6} under three categories --- example sources where the peak frequency is in agreement across all three datasets, example sources where the peak frequency for LCO agrees with that of ZTF DR5 but not with ZTF DR2, and example sources where the peak frequency across all three datasets are in disagreement. We note that with the exception of (21), the peak frequency obtained with DR2 and DR5 simultaneously agree only when it also agrees with the peak frequency obtained with LCO data. Agreement of our peaks with DR5 but not DR2 as well as the observed lower noise level with DR5 data compared with DR2 data, in all cases presented, highlights an improvement in the data quality in the newer ZTF data release.

We highlight two possible reasons for the higher number of detections obtained with our LCO observations compared to the ZTF Catalog of Variable Stars compiled by C20 using ZTF DR2 observations as well as compared to ZTF DR5.

\begin{figure*}[!ht]
    \centering
    \includegraphics[width=0.7\textwidth]{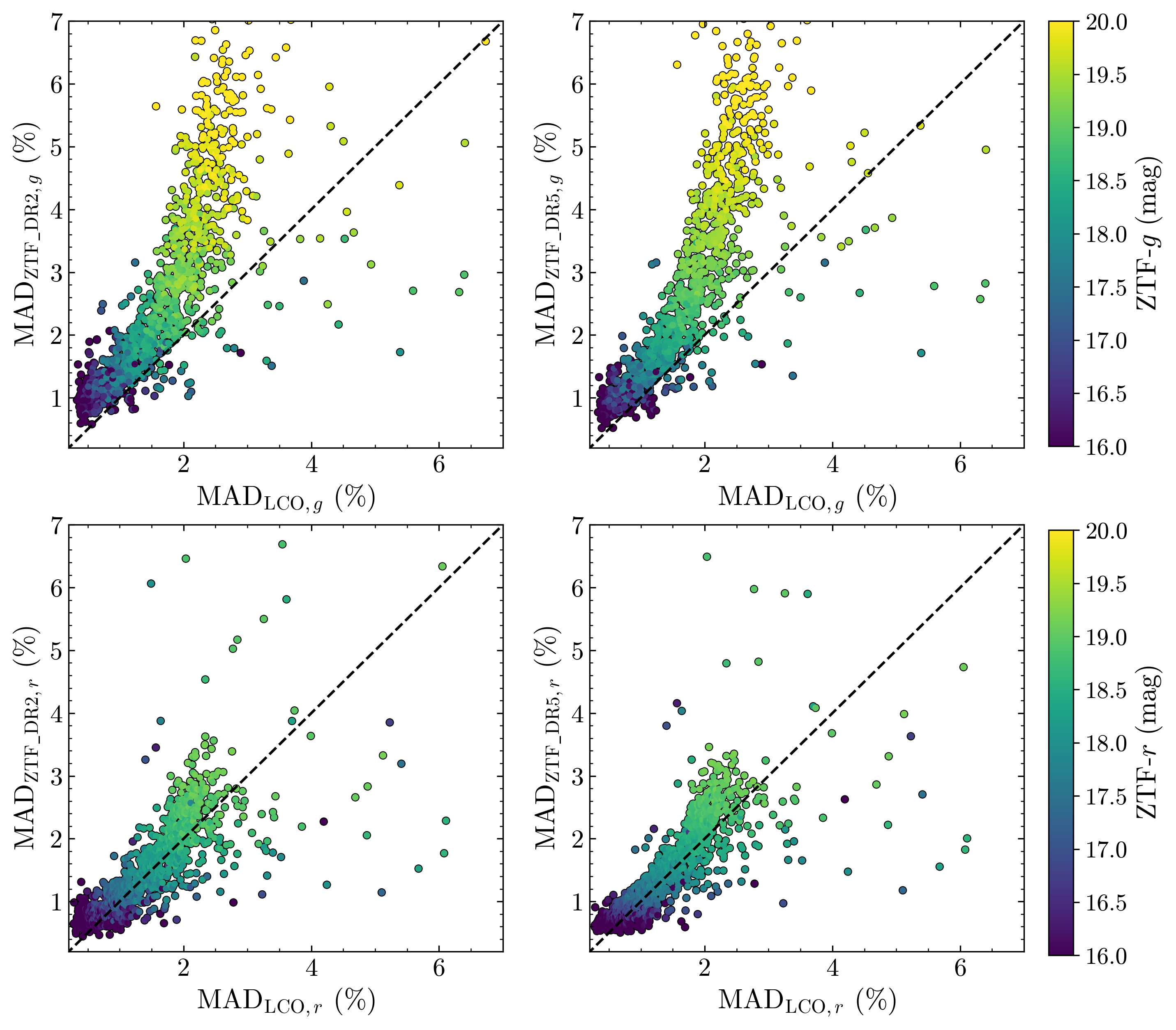}
    \caption{Comparison of ZTF and LCO photometry. Each panel compares the median absolute deviations (MAD) of the $g$-band ({\em top panels}) and $r$-band ({\em bottom panels}) light curves of LCO with those of ZTF DR2 ({\em left panels}) and DR5 ({\em right panels}). A diagonal line in each panel represents a 1:1 relationship. The data points are color coded by their median ZTF DR5 magnitudes. All variable sources identified in this work and the white dwarf with transiting planetary debris -- ZTF\,J0139+5245 -- are excluded from these figures. In general our LCO photometry exhibits improved photometric precision per-exposure compared to ZTF, especially for faint sources in the $g$-band, likely due to our longer exposure times of 2$-$3 minutes compared to 30 seconds for ZTF. Since we used the PS1 $r$-band to perform a magnitude cut at 19.25 mag, the $r$-band magnitudes for our sample do not go nearly as faint as the $g$-band since our sources are mostly red. The outliers in these panels that have much higher MAD in LCO than ZTF are primarily sources found near to CCD edges and have unreliable photometry, or are faint sources contaminated by nearby bright stars.}
    \label{fig:7}
\end{figure*}

First, we quantify and compare the photometric precision per exposure for LCO and ZTF photometry using the Median Absolute Deviation (MAD) indicator. Figure~\ref{fig:7} compares the MAD of $g$ and $r$-band light curves obtained with LCO with those obtained with ZTF DR2 and DR5. We exclude all variable sources identified in this work along with the white dwarf with transiting planetary debris --- ZTF\,J0139+5245. We observe that our LCO photometry has a lower MAD on average at essentially all magnitudes, and performs significantly better in the $g$-band at faint magnitudes. In the $r$-band, the LCO photometric precision is comparable to ZTF for most of our objects, though we do not sample the same range of magnitudes as in the $g$-band due to our imposed magnitude limits (see Section~\ref{2.2}). Aside from the main groups of objects in Figure~\ref{fig:7}, we also find several outliers with much higher MAD values in LCO than ZTF. We investigated the light curves and LCO images of these objects and found that they all fall into one of two categories which make their LCO photometry unreliable: (1) sources which are frequently found near to or on the edge of the CCD, and (2) faint sources contaminated by nearby bright stars. These outliers represent a small subset of our total sample, but indicate that the source selection process we used in Section~\ref{2.2} is not perfect at removing potentially problematic sources. While several factors contribute to the photometric precision for a given set of observations, such as atmospheric transmission, CCD quantum efficiency, telescope aperture size, and many others, we suspect that a major factor improving the photometric precision of our LCO observations compared to ZTF observations is the difference in exposure times. Our LCO exposure times range from 2$-$3 minutes while the ZTF exposure times are always 30 seconds.

Second, we find that the photometric time-sampling of our observations with LCO is greater than both ZTF DR5 and DR2 by about a factor of 2.5 --- we have an average of 1141 photometric datapoints across a time baseline of 537 days with LCO ($\approx$ 2.12 per day), 553 photometric datapoints across a time baseline of 719 days with ZTF DR5 ($\approx$ 0.77 per day), and 293 photometric datapoints across a time baseline of 368 days with ZTF DR2 ($\approx$ 0.79 per day).

The result of a longer exposure time and denser photometric time sampling with the LCO dataset is lower noise levels in the periodograms compared to ZTF. We quantify the noise levels using the statistical mean of the periodogram data. We find that, on average, the ZTF noise level is $\approx$ 3.6 times greater than the LCO noise level. Consequently, this implies that a 3$\sigma$ peak detection in the LCO periodograms would correspond to a 0.8$\sigma$ non-detection in the ZTF periodograms, assuming similar peak amplitudes. Regardless of the exact reason, however, the demonstrated improvement in photometric precision and noise levels enables us to detect lower-amplitude variability in the objects covered by our LCO images.

\subsection{ATLAS Catalog of Variable Stars}
We performed a box search of size 25 arcminutes on our FOV centered on the object J013906.17+524536.8 in the ATLAS catalog of variable stars provided by \citet[hereafter H18]{2018AJ....156..241H}.
We then cross-matched our list of candidate variable stars with the results of the box search. In our image's FOV, H18 have identified 84 variable sources. We note that they classify 77/84 of the detections as dubious sources, defined as stars that might not be real variables. Our candidate variable star list confirms 6/7 of the remaining sources that were assigned a classifications other than dubious. Our investigation revealed that our crowding limitation selection criterion eliminated one of these 7 sources from consideration. Moreover, we have confirmed variability in two additional matches that were classified as dubious by H18. Out of the remaining 75 dubious sources, 44 sources made it through our selection criterion for visual inspection. We found no variability in these objects. The remaining 31 dubious sources have been excluded from consideration by our selection criterion and thus their variability can neither be confirmed nor denied.

The 6/7 sources in H18 with classifications other than dubious matched with
(31), (32), (33), (34), (36), and (38) in our candidate variable star list. We confirmed variability in source (7), classified as a BY Draconis variable, and (29), classified as an EA type eclipsing binary system, which were classified as dubious sources by H18.

H18 classify (32), (34), and (36) as SINE variables with periods of 0.257171d, 0.332667d, and 0.627621d respectively. SINE variables are defined as sinusoidal variables, likely dominated by ellipsoidal variables, in which stars exhibit simple sine-wave variability with little residual noise. Our classifications and period identifications for the above sources are in agreement, with the exception of source (34) for which we identified secondary eclipses and hence a period of 0.665d. While H18 broadly classify the above sources as SINE variables based on features identified using machine-learning algorithms, a technique very well-suited to large sample sizes, given our small sample size, we are able to classify these objects more specifically as Eclipsing Binaries of type EW.

H18 classify (31) as a CBH with a period of 0.422072d. CBH stands for close binary, half period. These stars are contact or near-contact eclipsing binaries for which the Fourier fit has settled on half the correct period and hence has overlapped the primary and secondary eclipses. They also classify (33) as an IRR with a period of 0.382738d, where IRR stands for ``irregular" variables. This class serves as a ``catch-all'' bin for objects that do not seem to fit into any of their more specific categories. Most of the stars classified as IRR do not show coherent variations that can be folded cleanly with a single period.

Our identified periods for the above sources are in agreement but our classification for (33) is in disagreement. Based on the light curve presented in Figure~\ref{fig:5}, we classify (33) as an Eclipsing Binary of type EW.

\section{Conclusion}
\label{conclusion}
In this paper, we have presented the results of aperture photometry and Fourier analysis performed on 2227 sources within 1560 LCO images spanning 537 days and cross-matched with the PS1 catalogue. Our main findings are as follows:

1. We have identified 38 variable sources and have classified them on the basis of characteristic light curve properties, periods, as well as their locations within the {\em Gaia} CMD: 28 candidate BY Draconis variables, 3 candidate eclipsing binaries of type EA, and 7 candidate eclipsing binaries of type EW. The complete results are presented in Table \ref{table} and Figures~\ref{fig:4} and \ref{fig:5}. In assigning preliminary classifications to the detected sources, we have demonstrated the applicability of the {\em Gaia} CMD as a powerful classification tool for future surveys to utilize.

2. We determined that 27 of the 38 detected variable sources are newly discovered or newly classified (71\%) based on comparisons with previously published catalogs, thereby increasing the number of detections in the field-of-view under consideration by a factor of $\approx$ 2.5. In addition, we confirmed variability in two additional sources previously observed but classified as dubious sources.

3. Using the median absolute standard deviations (MAD) of the $g$-band and $r$-band light curves obtained with LCO and ZTF photometric data, we demonstrate that, in general, our LCO photometry exhibits improved photometric precision per-exposure compared to ZTF, especially for faint sources in $g$-band. We attribute this improvement to LCO's longer exposure times of 2$-$3 minutes compared to the 30 seconds of ZTF. The greater time-sampling of LCO photometry combined with the longer exposure time reduces the noise levels in our periodograms and improves our sensitivity to potential periodicities by a factor of $\approx$ 3.6. This enables us to increase the total number of detections in this field-of-view.

4. Finally, since the majority of sources we detect do not have spectral classifications, we note that it may be beneficial to conduct follow-up spectroscopic observations to further solidify the variable star classifications we provide. Further, additional time-series photometry for objects with the least significant periodogram peaks may enable better characterization of these sources.

\section*{Acknowledgements}
This work was conducted as part of the White Dwarf Stars Research Stream \citep{Montgomery10c,Winget13} in the Freshman Research Initiative (FRI) program
\citep{Clark16,Rodenbusch17} in the College of Natural Sciences at the University of Texas at Austin. A.S. acknowledges financial support from the Freshman Research Initiative Summer Fellowship awarded by the College of Natural Sciences at the University of Texas at Austin.

Z.P.V. and M.H.M. acknowledge support from the United States Department of Energy under grant DE-SC0010623, the National Science Foundation under grant AST-1707419, and the Wootton Center for Astrophysical Plasma Properties under the United States Department of Energy collaborative agreement DE-NA0003843. M.H.M. acknowledges support from the NASA ADAP program under grant 80NSSC20K0455.

This work is based on observations obtained with the Samuel Oschin 48-inch Telescope at the Palomar Observatory as part of the Zwicky Transient Facility project, which is supported by NSF grant No. AST-1440341 and the participating institutions of the ZTF collaboration. We make use of observations and reduction software from the LCO network, and include data from the Pan-STARRS1 Surveys (PS1) and the PS1 public science archive, which have been made possible through contributions by participating institutions of the PS1 collaboration (https://panstarrs.stsci.edu/). We also make use of data from the European Space Agency (ESA) mission {\em Gaia} (https://www.cosmos.esa.int/gaia) processed by the {\em Gaia} Data Processing and Analysis Consortium (DPAC, https://www.cosmos.esa.int/web/gaia/dpac/consortium).

\emph{Additional Software/Resources}: {\sc Astropy} \citep{2013A&A...558A..33A, 2018AJ....156..123A}, {\sc Photoutils} \citep{2019zndo...3568287B}, {\sc Pyriod} maintained by Dr.\ Keaton Bell \citep{2020AAS...23510606B}, the NASA Astrophysics Data System (ADS), and SIMBAD and VizieR (operated at CDS, Strasbourg, France).

\bibliography{\string Sanghi_Astro_Bib}

\end{document}